\documentclass[11pt]{aa}  

\usepackage{graphicx}
\usepackage{txfonts}
\usepackage{soul}
\usepackage{float}
\usepackage[colorlinks,citecolor=blue,urlcolor=blue]{hyperref}

\begin{document} 

    \title{Umbral oscillations in the photosphere}
    \subtitle{A comprehensive statistical study}

    \author{M. Berretti \inst{1,2}
            \and
            M. Stangalini \inst{3}
            \and
            G. Verth \inst{4}
            \and
            V. Fedun \inst{5}
            \and
            S. Jafarzadeh \inst{6,7}
            \and
            D.~B. Jess \inst{6,8}
            \and
            F. Berrilli \inst{2}
            }

    \authorrunning{Berretti et al.}

   \institute{University of Trento,
              Via Calepina 14, 38122 Trento, Italy,
              \email{michele.berretti@unitn.it}
         \and
             University of Rome Tor Vergata, Department of Physics, Via della Ricerca Scientifica 3, 00133 Rome, Italy
         \and 
             ASI Italian Space Agency, Via del Politecnico snc, 00133 Rome, Italy,
             \email{marco.stangalini@asi.it}
          \and
            Plasma Dynamics Group, School of Mathematical and Physical Sciences, The University of Sheffield, Hicks Building, Hounsfield Road, Sheffield, S3 7RH, UK
          \and
            Plasma Dynamics Group, School of Electrical and Electronic Engineering, The University of Sheffield, Mappin Street, Sheffield, S1 3JD, UK
          \and
             Astrophysics Research Centre, School of Mathematics and Physics, Queen’s University Belfast, Belfast, BT7 1NN, Northern Ireland, UK
          \and
          Niels Bohr International Academy, Niels Bohr Institute, Blegdamsvej 17, DK-2100 Copenhagen, Denmark
            \and
            Department of Physics and Astronomy, California State University Northridge, Northridge, CA 91330, USA}

   \date{}

  \abstract
  {It is well-known that the global acoustic oscillations of the Sun's atmosphere can excite resonance modes within large-scale magnetic concentrations. These structures are conduits of energy between the different layers of the solar atmosphere, and understanding their dynamics can explain the processes behind coronal heating and solar wind acceleration. In this work, we studied the Doppler velocity spectrum of more than a thousand large-scale magnetic structures (i.e., sunspots) in the solar photosphere that crossed near the disk centre of the Sun. We exploited the excellent stability and seeing-free conditions of the Helioseismic and Magnetic Imager (HMI) instrument onboard the Solar Dynamics Observatory (SDO) to cover nearly seven years of observations, providing the most comprehensive statistical analysis of its kind. Here, we show that the power spectra of the umbra of sunspots in the photosphere is remarkably different from the one of quiet-Sun regions, with both exhibiting a primary peak at 3.3 mHz, but the sunspot umbrae also displaying a closely packed series of secondary peaks in the $4-6$~mHz band. Understanding the origin of such peaks is a challenging task. Here, we explore several possible explanations for the observed oscillations, all pointing toward a potential resonant interaction within these structures and an unknown driver. Our observational result provides further insight into the magnetic connectivity between the different layers of the dynamic atmosphere of the Sun.}

               
   \maketitle

    \section{Introduction}

    The general understanding of oscillations in the Sun's atmosphere is that the photosphere is dominated by the global resonant modes of the entire stellar structure at $5$~min (i.e., $3$~mHz), while, moving upwards to the chromosphere, the dominant period shifts to $3$~min ($5$~mHz) (see earlier reviews by \cite{cram_physics_1981, lites_sunspot_1992, bogdan_sunspot_2000} and more recent works by \cite{khomenko_oscillations_2015, jess_waves_2023}). This behaviour is considered to be the result of an acoustic cut-off caused by the stratification of the solar atmosphere \citep[][]{wisniewska_observational_2016, felipe_height_2018, felipe_numerical_2020}. Large-scale magnetic concentrations, such as sunspots and pores, are embedded in this environment and serve as conduits between the layers of the solar atmosphere \citep[][]{giovanelli_motions_1978, lites_sunspot_1985}. These structures are thus crucial to the dynamics of magnetically dominated solar atmospheres and are possible candidates to explain coronal heating \citep[][]{goossens_energy_2013, van_doorsselaere_energy_2014, van_doorsselaere_coronal_2020}, solar wind acceleration \citep[][]{de_pontieu_chromospheric_2007}, and changes in plasma composition throughout the solar atmosphere \citep{2021ApJ...907...16B, 2021RSPTA.37900216S, 2021A&A...656A..87M, 2024PhRvL.132u5201M}.

    Since their first observation in \cite{beckers_chromospheric_1969}, oscillations within the umbra of sunspots have been thoroughly studied. In \cite{edwin_wave_1983} the authors predicted the excitation of resonance modes inside these structures, and after that many authors reported observational evidence of a plethora of wave modes hosted by sunspots \citep[][to name but a few]{centeno_oscillations_2006, dorotovic_identification_2008, jess_alfven_2009, morton_observations_2011, morton_evidence_2013, keys_photospheric_2018, stangalini_propagating_2018, albidah_proper_2021}. In this regard, in \cite{albidah_magnetohydrodynamic_2022}, the authors used Proper Orthogonal Decomposition \citep[POD; a similar approach to Principal Component Analysis, PCA;][]{pearson_lines_1901} and Dynamic Mode Decomposition \citep[DMD;][]{schmid_dynamic_2010} techniques to study the wave modes in selected sunspots observed with the HARDcam instrument \citep[][]{jess_source_2012} at the National Solar Observatory's Dunn Solar Telescope (DST; New Mexico, USA). POD and DMD are both model order reduction techniques that are used to reduce computational complexity by reducing the degrees of freedom, simplifying their state space without loss of information. The authors provided evidence of multiple high-order MHD modes inside the observed structures. Recently, in \cite{jafarzadeh_sausage_2024} the authors reported the concurrent excitation of multiple-order MHD modes in solar magnetic pores observed with the High Resolution Telescope \citep[HRT,][]{gandorfer_high_2018} of the Polarimetric and Helioseismic Imager \citep[PHI;][]{solanki_polarimetric_2020} on board of Solar Orbiter \citep[SO;][]{muller_solar_2020}.

    \begin{figure*}[t!]
       \centering
       \includegraphics[width=\hsize]{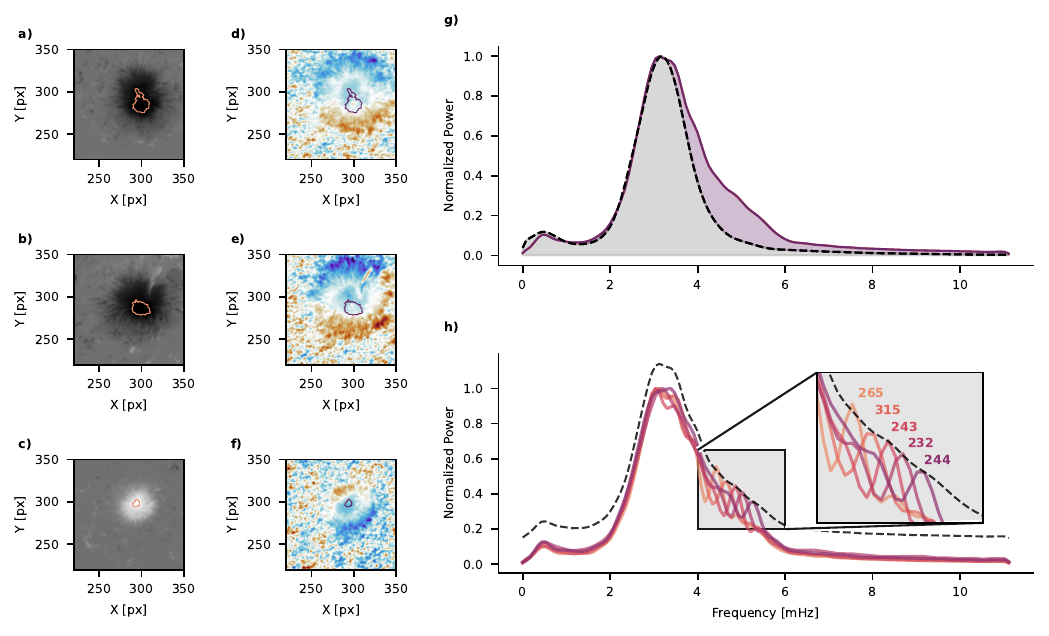}
       \caption{\textbf{Doppler velocity analysis of sunspots.} \textbf{a-c:} Magnetograms of three randomly selected sunspots. The orange contours mark the region in which the magnetic field is above $1800$~G, considered to be just about the umbra-penumbra boundary \citep{jurcak_magnetic_2018}. All the magnetograms are saturated between $-1800$~G and $1800$~G \textbf{d-f:} Corresponding Dopplergrams of the same sunspots. The violet contours indicate the umbral region, inherited by the magnetograms. \textbf{g:} Mean Doppler-velocity power spectrum for all sunspots (violet; solid line) and for quiet-Sun regions (grey; dashed line). \textbf{h:} Mean power spectrum of the sunspots classified by the positions of the frequency of their secondary peaks. Following the order in the legend, the classes are centred respectively at $4.3,\, 4.5,\,4.7,\,4.9,\,5.1$~mHz, with a width of $0.3$~mHz. The number of sunspots per class is indicated above each peak. The black dashed line represents the mean power spectrum shown in violet in panel g, shifted upwards by $0.14$ to aid visualisation.}
       \label{fig:1}
   \end{figure*}

     Suggested by \cite{thomas_umbral_1982}, and following the findings by \cite{moore_generation_1973} and \cite{mullan_can_1973}, umbral oscillations are thought to be the resonant response to $p$-modes in subphotospheric convective layers of the Sun's atmosphere. In \cite{spruit_conversion_1992}, the authors studied the interaction of acoustic oscillation and large-scale magnetic structures in a simple model atmosphere, described as a plane-parallel stratified adiabatic polytrope with an embedded vertical magnetic structure. They studied the conversion of $p$-modes to slow modes and found evidence of absorption of these modes by the magnetic structures. In \cite{dsilva_acoustic_1994}, the author showed that sunspots, beyond $p$-mode absorption and conversion, are very efficient at mixing modes, with a high sensitivity to the morphological properties of the magnetic structure. This led to the idea that large magnetic structures absorb the $p$-modes from the nearby environment and convert purely acoustic oscillations to magneto-acoustic wave modes of similar frequency.

    Although most of the studies have reported $5$-minute ($3$ mHz) umbral oscillations in the photosphere, some authors also reported 3-minute oscillations  \citep{beckers_oscillatory_1972, rice_oscillatory_1973, soltau_velocity_1976, schroeter_time_1976}. In \cite{abdelatif_interaction_1986}, the authors observed a distinct 3-minute oscillation in the line-of-sight (LOS) velocity of two sunspots concentrated in the darkest regions of the umbrae. Using spectral data from the Fast Imaging Solar Spectrograph \citep[FISS;][]{chae_fast_2013}, \cite{chae_photospheric_2017} investigated the origin of these 3-minute oscillations in the photosphere. They concluded, despite the limitations on the resolution of the instrument, that magnetoconvection occurring in lightbridges and umbral dots may generate the observed periodicity. In \cite{stangalini_novel_2021} the authors studied a peculiar magnetic pore, observed with the Interferometric Bidimensional Spectropolarimeter \citep[IBIS;][]{cavallini_ibis_2006} at the DST, showing Doppler velocity oscillations with a dominant frequency of 5~mHz (i.e., a period of 3~min) in the solar photosphere, and no sight of 5-minute oscillations typical of the photospheric driver. Using the same instrument, in \cite{stangalini_large_2022} the authors studied an unusually intense sunspot observed for three hours. They used the $B-\omega$ diagram \citep{stangalini_novel_2021} to distinguish the spectra inside and outside the magnetic structure, POD to identify the MHD modes, and phase-lag analysis to infer the phase speed of the observed oscillations. Their results strengthened the idea that the power spectra of large-scale magnetic structures differ greatly from those of the environment in which they are embedded, with frequency peaks other than the usual dominant peak of the p-modes.

    In this work, we analysed the LOS velocity spectrum of nearly two thousand large-scale magnetic concentrations that crossed near the disk centre of the Sun from April 2011 to December 2017. We used the excellent stability of the Helioseismic and Magnetic Imager \citep[HMI;][]{schou_design_2012}, the magnetometer onboard of NASA's Solar Dynamics Observatory \citep[SDO;][]{pesnell_solar_2012}, to provide the most comprehensive statistical analysis of the dynamics of sunspots in the photosphere. In particular, we find that for the majority of the sunspots included in this work, the spectrum of their umbra is significantly in contrast to that of the surrounding quiet-Sun regions, providing further insights into the magnetic connectivity of the solar atmosphere.

    \section{Dataset and analysis}

    The sunspots analysed in this work were observed with SDO/HMI in the Fe~{\sc i}~$617.3$~nm absorption line, with a cadence of $45$ seconds in monthly observation windows of $45$ minutes. We restricted our analysis to large-scale magnetic structures (i.e., sunspots and pores with an average umbral diameter of around 3000~km) that crossed within $0.4$ solar radii of the disk centre. The images were co-registered to remove the contribution of the solar rotation. The data set consists of nearly seven years of observations, including the LOS velocity and magnetic field of the selected sunspots from the Debrecen Photoheliographic Data catalogue \citep[][]{baranyi_-line_2016, gyori_comparative_2017}. Each Dopplergram is constructed using six filtergrams for each of the six positions in the Fe~{\sc i}~$617.3$~nm spectral line, resulting in a formation height of approximately $100$~Km \citep{fleck_formation_2011}. In total, the data set contains $612$ active regions and a total of $1714$ unique sunspots and pores. 

    In the leftmost panels of Fig. \ref{fig:1} (panels a--f), we show the magnetogram (left) and the corresponding Dopplergram (right) for three randomly selected sunspots in the dataset. The contours highlight the region of sunspots with magnetic field values above 1800~G, corresponding to just about the umbra-penumbra boundary, as reported by \cite{jurcak_magnetic_2018}. The regions of interest (RoI) were selected using the sunspots routine of the Solar Feature Tracking code (SoFT; \citealt{Berretti2024SoFT}). In this case, SoFT acts as a simple threshold discriminator, assigning a label to clumps with magnetic fields greater than 1800~G, allowing one to distinguish between the different magnetic structures in an active region. Of the 1714 sunspots included in this work, 759 met the magnetic field threshold and were selected for our analysis.

    To obtain the power spectra of the LOS velocity oscillation with the Fast Fourier Transform (FFT), we averaged the power spectra of all individual pixels within the umbra. Prior to computing these power spectra, to mitigate the effect of bad pixels hijacking the mean, we first applied a boxcar smoothing with a kernel of $3\times3$ pixels. Additionally, to ensure the reliability of the power spectral densities obtained, we followed the procedure described in \cite{jess_waves_2023}. First, we removed any linear trend, potentially linked to slow evolutions or residual solar rotation contributions that survived after the coalignment, by subtracting a linear least-squares fit. To increase the display resolution of our spectra and mitigate the effects of the truncation of the time-series, respectively, we zero-padded the series to a length of 256 and applied a Tukey window function with an alpha parameter of 0.1 (apodising). Finally, each sunspot's mean power spectrum was calculated and normalised to its maximum value, facilitating subsequent comparison across all sunspots.

   \begin{figure}[h!]
       \centering
       \includegraphics[width=\linewidth]{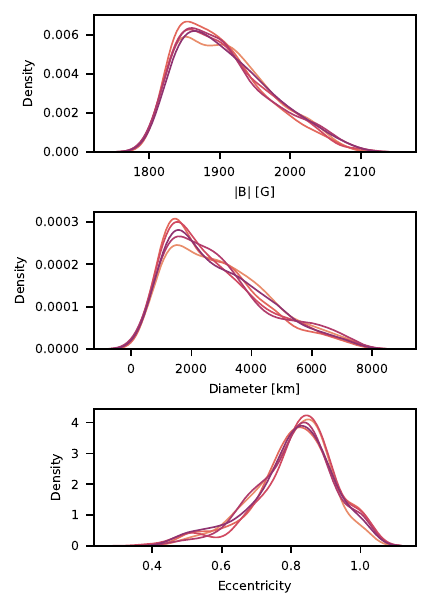}
       \caption{Statistical distributions of the magnetic field (top) and equivalent diameter (middle) and eccentricity (bottom) of the different classes of sunspots. The colours are inherited from Fig. \ref{fig:1} for each class. The equivalent diameters were estimated considering the umbrae as perfectly round. The eccentricity is defined as the filling factor of the magnetic feature with respect to a circle with a radius corresponding to half of the equivalent diameter of the feature and centred in the barycentre of it.}
       \label{fig:2}
   \end{figure}

    \section{Results}

    In Fig. \ref{fig:1}g we show in violet the mean spectra obtained by averaging the normalised power spectrum of the $759$ sunspots included in this work. In contrast, the black dashed line represents the mean spectrum of $p$-modes, obtained by averaging the $300$ spectra of the $50\times50$~ pixels of the quiet Sun within the field of view. We can see a slight broadening of the peak at $3.3$~mHz and a significant bump in the $4-6$~mHz band of the sunspots' power spectra. 

    We can now filter the power spectra of the sunspots based on the positions of their secondary peaks. To this end, we selected five different overlapping frequency bands, starting from $4.3$~mHz (as the centre of the first window) with a distance of $0.2$~mHz from each other and $0.3$~mHz width. In Fig. \ref{fig:1}h, we show the mean spectra of each of the sunspot classes. Each class comprises around $300$ spectra, as shown in the coloured text at the bottom panel of Fig.~\ref{fig:1}, but a single spectrum can belong to multiple classes. It should be noted that only $88$ out of $759$ have no power in any of these bands and, on the other hand, exhibit only a peak at $3.3$~mHz. 

    Finally, in Fig. \ref{fig:2} we show the statistical distributions of the magnetic field (top), the equivalent diameter (middle) and eccentricity (bottom) of the 5 selected classes, maintaining the same colour coding as in Fig.~\ref{fig:1}.  In place of the histograms, we opted for the probability density curves (KDE). Indeed, histograms approximate the underlying probability distributions by binning the data and counting the occurrences in each bin, making them a distorted estimator of the actual probability distribution. In contrast, KDE tries to smooth the approximation performed by the histogram using a kernel function, producing a continuous density estimate. The choice of the kernel function and, mostly, its bandwidth becomes critical. Here, we used a normal kernel using Scott's rule \citep{scott_optimal_1979} to infer the optimal value of the bandwidth. However, when working with data sets containing over-represented elements, the KDE can smooth and merge peaks visible in the histogram. Histograms, however, are reported in appendix~\ref{ap:1}. Finally, the observed statistical distributions are compatible with those shown in Fig. 3 of \cite{livingston_decreasing_2012}, with a similar asymmetry between the left and right wings of the distribution. 

    \section{Discussions}
    
    The enhancement in the average power spectrum seen in Fig.~\ref{fig:1}g is the result of the overlap of multiple classes of spectra with a frequency peak in small frequency intervals of $0.3$~mHz, starting from $4.3$~mHz to $5.1$~mHz. Understanding the origin of these peaks in the Doppler velocity spectrum of the $759$ sunspots is a challenging task. The existence of a frequency band centred at about $5$~mHz in the photosphere is not, in itself, a new finding. The strength of this study lies in the unprecedented statistical consistency of our results, showing multiple peaks in the spectra of sunspots' umbrae, proving that it is much different from that of the nearby quiet Sun. As a proper interpretation of the results presented in this work is, as of this date, impossible due to the lack of multi-height observations and simulations capable of coupling the dynamics of local plasma, magnetic structures and the global dynamo, we can only provide an interpretation of our results by revisiting the many theories, sometimes forgotten, that are already available in the past literature. 

    \begin{enumerate}
        \item One could argue that the peaks in the $4-6$~mHz band are the resonance frequencies given by the intrinsic response of the magnetic structures to subphotospheric convective flows. However, the lack of correlation between the position of the frequency peaks and key physical parameters such as the area, magnetic field, and shape of the umbra discourages this interpretation. In other words, one would expect that similar structures should have similar excited frequencies, but this is not the case given the heterogeneous sample. Therefore, we are prone to believe that the observed secondary peaks is the result of an underlying driver, acting on all the magnetic structures considered and forcing this secondary photospheric frequency centred at about $5$~mHz. However, establishing the nature of the driver and how it interacts with the magnetic concentrations considered in this work would require simulations of the entire photosphere and convection zone coupled with the solar dynamo -- an effort that, to date, is simply unachievable due to the computational resources required. 
        \item In the 1980s, a series of works by Thomas, Scheuer, and Zhugzhda attempted to explain the origin of the observed 3-minute oscillations in the umbra of sunspots in the photosphere. \cite{scheuer_umbral_1981} argued that such oscillations, following earlier work by \cite{uchida_oscillations_1975}, are driven by subphotospheric overstable convection, exciting these nearly trapped fast magneto-acoustic waves \citep{thomas_umbral_1982}. In \cite{centeno_oscillations_2006} the authors studied the velocity oscillations of sunspots in the photosphere and chromosphere and argued that the observed dominant frequency of $5$~mHz in the chromosphere is the result of the linear propagation of waves from the photosphere. On the other hand, the Zhugzhda model predicted the existence of a resonance cavity in the chromosphere driven by the acoustic noise of convection \citep[][]{zhugzhda_seismology_1983, zhugzhda_resonance_1984}. Indeed, in \cite{hollweg_new_1979} the authors first suggested the possibility of the existence of chromospheric resonances by solving the equations of plasma motion in a model atmosphere and demonstrated the occurrence of resonances between the driving fast mode and the driven acoustic oscillations. Further advances in the existence of a chromospheric cavity were made by \cite{botha_chromospheric_2011}, where the authors numerically showed that the dominant period of 3 min observed in the chromosphere can be supported by a resonance cavity without the need for a source in the photosphere. Finally, in \cite{jess_chromospheric_2020}, the authors reported strong observational evidence for the existence of a resonance cavity in the chromosphere. Ultimately, in \cite{lites_photoelectric_1984}, \cite{thomas_umbral_1984} and \cite{zhugzhda_resonance_1984}, the authors suggested the existence of multiple resonance cavities in the solar atmosphere, finally explaining the presence of multiple closely packed peaks near the $3-6$~mHz band. The two cavities are the photospheric one generated by the reflective layer below the photosphere owing to the increase in the sound speed and the one in the chromosphere due to the reflective layer arising from the increase in Alfvén velocity. Our results could very well fit this interpretation, although they cannot be intended as observational evidence for the existence of a photospheric cavity and further proof would be required.
        \item Alternatively, it is well known that regions of intense magnetic field inhibit heat transfer, leading to a localised lowering of temperature and opacity. This process results in an optical depth difference between the umbrae of sunspots and the neighbouring quiet Sun, referred to as the Wilson depression \citep[][]{wilson_observations_1774}. In \cite{loptien_connecting_2020}, the authors studied the relationship between the properties of magnetic structures and the depth of the Wilson depression using two methods: one based on the divergence of the magnetic field and the other on balancing the horizontal forces between the region inside and outside the sunspot. In their work, they found that the depth of the Wilson depression, for sunspots compatible with the ones in our work, can vary between $600$ km for the former method and $200$ km for the latter. The formation height of the Fe~{\sc i}~$617.3$~nm absorption line used by HMI was estimated to be around $100$ km by \cite{fleck_formation_2011}. It is also worth mentioning that the mixing of different heights in the HMI pipeline \citep{scherrer_helioseismic_2012} and the height dependence of the acoustic cutoff \citep{wisniewska_observational_2016} might also contribute to bringing the height at which our signal originates further closer together.
    \end{enumerate}
    
    Despite the possible interpretations of our results, the key aspect to underline is that the vast majority of sunspots that crossed near the centre disk show a secondary peak in the $4-6$~band, a frequency that is usually seen to be dominant in the chromosphere. This is in fact observational evidence, with unprecedented statical consistency, that the spectra of sunspots in the photosphere greatly differ from that of quiet-Sun regions, an effect observed in the past and predicted from the theory of the early 1980s but seemingly overlooked over the past two decades. The suggested explanations here are a tentative effort to provide different points of view on the challenging task of understanding the magnetic connection between the layers of the solar atmosphere. Lack of ({\sc i}) multiheight observations, ({\sc ii}) full-disk simulations of the solar photosphere capable of describing the interaction at global scales within magnetic structures, the global dynamo, and the convective envelope of the Sun, and ({\sc iii}) high-resolution instrumentation to reduce the effects of stray light on the observed umbrae, we cannot provide a definitive answer to the observed frequency peaks.
    
    Furthermore, in \cite{berretti_unexpected_2024}, the authors analysed horizontal oscillations of small-scale magnetic elements in the photosphere and found a consistent dominant frequency of approximately 5~mHz. In this study, we have examined oscillations in the LOS velocities within sunspot umbrae, which represent significantly larger magnetic structures. Despite the difference in scale and the nature of the observed oscillations (horizontal motions versus LOS velocities; transversal versus longitudinal waves), we also identify a frequency band centred around 5~mHz in these larger structures. However, this band is broader than the narrow peak found in the small-scale elements. Given the distinct characteristics of these two kinds of oscillations, we speculate that a global driver operating at around 5~mHz may be influencing both. Nevertheless, the findings of this work and those of \cite{berretti_unexpected_2024} reveal a prevalent 5~mHz oscillation in distinct solar features, suggesting the influence of one or more drivers linked to this frequency. Further investigation (using, e.g., advanced numerical simulations) is needed to determine the nature and origin of these drivers, and whether a single global mechanism or multiple independent sources are responsible.
    
    \section{Conclusions}

    In this work, we studied the spectra of Doppler velocity oscillations of the umbra of 759 sunspots in the photosphere observed with SDO/HMI. Our results show that the spectra of the selected structures differ greatly from those of quiet Sun regions, with a closely packed series of secondary peaks between $4$~mHz and $6$~mHz. Although a definitive answer on the origin of these peaks can be quite challenging to provide, in the Discussions section of this work, we have highlighted a few possible explanations that were extensively discussed in the past decades. Future works will require tomographic observations of sunspots and high-resolution instrumentation to study the magnetic connections between the different layers of the solar atmosphere.

    \begin{acknowledgements}
     MB acknowledges that this publication (communication/thesis/article, etc.) was produced while attending the PhD program in  PhD in Space Science and Technology at the University of Trento, Cycle XXXIX, with the support of a scholarship financed by the Ministerial Decree no. 118 of 2nd March 2023, based on the NRRP - funded by the European Union - NextGenerationEU - Mission 4 "Education and Research", Component 1 "Enhancement of the offer of educational services: from nurseries to universities” - Investment 4.1 “Extension of the number of research doctorates and innovative doctorates for public administration and cultural heritage” - CUP E66E23000110001.
    VF and GV are grateful to the Science and Technology Facilities Council (STFC) grants ST/V000977/1 and ST/Y001532/1. They also thank the Institute for Space-Earth Environmental Research (ISEE, International Joint Research Program, Nagoya University, Japan), the Royal Society, International Exchanges Scheme, collaboration with Greece (IES/R1/221095), India (IES/R1/211123) and Australia (IES/R3/213012) for the support provided.
    DBJ acknowledge support from the Leverhulme Trust via the Research Project Grant RPG-2019-371. DBJ wish to thank the UK Science and Technology Facilities Council (STFC) for the consolidated grants ST/T00021X/1 and ST/X000923/1. DBJ and SDTG also acknowledge funding from the UK Space Agency via the National Space Technology Programme (grant SSc-009).
    SJ acknowledges support from the Rosseland Centre for Solar Physics (RoCS), University of Oslo, Norway.
    We wish to acknowledge scientific discussions with the Waves in the Lower Solar Atmosphere (WaLSA; \href{https://WaLSA.team}{www.WaLSA.team}) team, which has been supported by the Research Council of Norway (project no. 262622), The Royal Society (award no. Hooke18b/SCTM; \citealt{2021RSPTA.37900169J}), and the International Space Science Institute (ISSI Team 502).

    \end{acknowledgements}

\bibliographystyle{aa}
\bibliography{main}

\begin{appendix}
\onecolumn
\section{Statistical distributions of the sunspots}\label{ap:1}

    \begin{figure}[h!]
       \centering
       \includegraphics[]{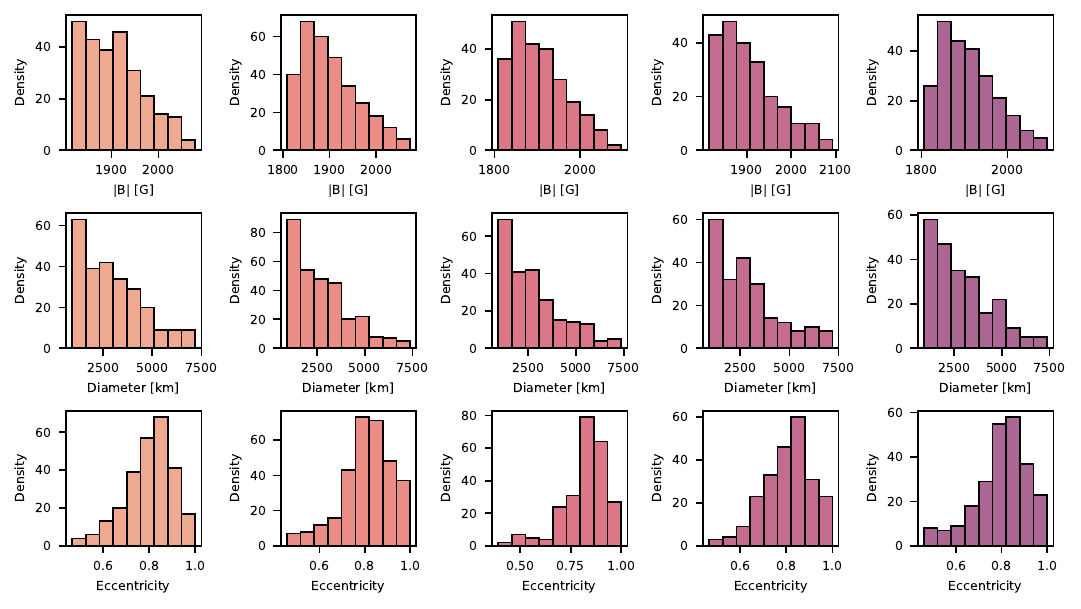}
       \caption{Statistical distributions of the average magnetic field of the umbrae (top row), equivalent diameter (middle row) and eccentricity (bottom row). The colours are inherited from the groups in Fig.~\ref{fig:1}.}
       \label{fig:3}
    \end{figure}
    
    In Fig.~\ref{fig:3}, we show the histograms of the statistical distributions shown in Fig.~\ref{fig:2}. The bin number and width were chosen according to the Scott rule \citep[see e.g.][]{scott_optimal_1979}. We wish to stress that these distributions are intended to characterise the population of sunspots considered in this work and are not intended to be representative of the sunspots in the Sun. The top row shows the statistical distribution of the mean magnetic field of the selected umbrae, and the middle row shows the statistical distributions of the equivalent diameter. Finally, the bottom one shows the distribution of the eccentricity of the umbrae. As mentioned in the main body of the manuscript, the observed statistical distributions for each group of sunspots are similar to each other, with no notable feature able to differentiate the umbrae in each group.

\end{appendix}

\end{document}